# Vibrational heat capacity of carbon nanotubes in low and ultra-low temperature regions

M. V. Avramenko, I. Yu. Golushko, A. E. Myasnikova, and S. B. Rochal

*Faculty of Physics, Southern Federal University, 5 Zorge Street, 344090 Rostov-on-Don, Russia*

We develop a continuous theory of low-frequency dynamics for single-walled carbon nanotubes (SWCNTs) weakly interacting with the environment. In the frame of the approach proposed we obtain temperature dependence of SWCNTs specific heat in the low ($T<40$ K) and ultra-low ($T<2.5$ K) temperature ranges. We take into account the main term in the coupling between SWCNT and the environment that slightly increases the frequencies of those SWCNT modes, which possess predominantly radial polarization. The coupling drastically decreases the density of phonon states in the lowest frequencies region. The theoretically predicted fall of the specific heat in the interval $T<2.5$ K properly explains available experimental data in contrast to the preceding approaches. The theory proposed can be the basis for studies of low-temperature heat capacity and phonon dynamics of many other single-walled and multi-walled tubular structures (boron nitride, transition metal dichalcogenides) which have emerged in the past decade.

## I. Introduction

Since their discovery [1], carbon nanotubes (CNTs) are holding the researchers attention because of high potential for various technological applications [2-4] due to their unique physical properties. Low temperature (LT) heat capacity is one of such important characteristics, and its temperature dependence is still a matter of debates. Unfortunately, there are only a few experimental papers presenting results of SWCNTs specific heat measurements. For the first time specific heat temperature dependences were obtained in the ranges $0.6<T<210$ K [5] and $2<T<300$ K [6]. Later, some experiments were carried out at ultra-low temperatures (ULT) and the obtained results differed significantly: by 2-4 times at $T<3.5$ K and 0.2-0.4 times at $3.5<T<4.5$ K.

It is generally agreed that the main contribution to heat capacity of CNTs and their bundles is made by phonon modes even at ULT. The theory of CNTs vibrational heat capacity has been developing rapidly. Currently, the two groups of approaches to theoretical studies of CNTs vibrational heat capacity exist. The first one include the methods based on computational modeling of SWCNTs vibrational dynamics, in particular, force-constant models [9-12] and methods of molecular dynamics [13, 14]. Large number of atoms in CNTs leads to huge amount of computation. Besides, complex multiparticle interaction potentials are to be used in order to yield satisfactory results. The authors of theoretical papers [9, 11, 12-14] state that their results are in quite good quantitative agreement with experimental data [6, 7]. However, in [9] it is reported that specific heat temperature dependence $T^x$ has two regimes: linear at LT, then quadratic. On the contrary, in [14] it is stated that $T^x$ function is quadratic at LT and then linear. Moreover, model [11] predicts the existence of three regimes instead of two in the temperature range $0<T<100$ K: $x=1/2$ at ULT (approximately at 0.5-0.8 K), then $x=1$ at $T<5$ K and $x>1$ at $T>5$ K. Slight adjustments to the results of [11] were made in [12] by specifying that $0.6<x<1$ in the second regime. Nevertheless, the specific heat temperature dependence obtained in [12] is in good agreement with experimental data [6] only in the range $2<T<30$ K.

Another well-known group of methods for SWCNTs thermal characteristics calculation is based on different continuous models, which usually consider SWCNT as a thin cylindrical shell [15-17]. Since continuous approaches do not take into account the discrete atomic structure of SWCNTs and therefore do not describe high-frequency range of their vibrational spectrum, it is impossible to obtain high-temperature heat capacity by means of these models. However, LT heat capacity of SWCNTs is entirely determined by their low-frequency dynamics, which, in turn, is almost independent of SWCNTs discrete structure [15, 18]. Thus LT heat capacities calculated in the frame of discrete and continuous models should be the same. Nevertheless, specific heat temperature dependence obtained in [16] is almost linear and, as a result, poorly describes experimental data [6]. In the revised version [17] of theory [16] it is demonstrated that specific heat temperature dependence has two regimes: $x=1$ at LT and $x>1$ at temperatures higher than 180-200 K.

Thus, up-to-date notions of SWCNTs LT heat capacity are quite contradictory and, obviously, that it is due to shortcomings of the models describing SWCNTs vibrational dynamics. In the recently published paper [18] low-frequency vibrational spectra of both single-walled and double-walled CNTs were successfully described in the frame of new continuous theory of two-dimensional membrane dynamics. Following the classical theory of elasticity [19], the preceding approaches related membrane topological elasticity to its macroscopic thickness. Such approximations are obviously incorrect for nanomembranes of one atom thickness (like graphene and SWCNTs) and, as it is shown below, lead to certain mistakes in describing low-frequency spectrum of SWCNTs. The theory [18] considers bending elasticity of membrane of one atom thickness as phenomenological characteristics and describes 2D membrane using only three independent elastic constants. Unfortunately, the approach [18] is valid for individual (free) nanotubes only. However, till now it is impossible to measure heat capacity of a single SWCNT. To understand the existing experimental data on SWCNTs bundles we develop approach [18] and take into account the weak interaction of the single SWCNT with the environment. We prove that this interaction influence on the heat capacity increases while the temperature decreases. Taking into account of this fact allows constructing a theory of SWCNTs heat capacity, which shows better agreement with experimental data [5-8] than the preceding models.

The paper is organized as follows. The second section is devoted to the analysis of SWCNTs dynamics, both individual and coupled with medium. In the third section we discuss LT heat capacity of SWCNTs and compare the results obtained with available experimental and theoretical data. The Appendix considers some existing continuous models, which describe the phonon dynamics of SWCNTs.

## II. Dynamics of SWCNTs interacting with the environment

In this section we develop approach [18], which considers graphene and nanotubes as 2D membranes without macroscopic thickness. Till now it is impossible to measure heat capacity of individual SWCNT, thus the papers [5-8] present experimental studies of specific heat of SWCNTs bundles. Obviously, in this case CNTs cannot be considered as individual ones, and environment makes some contribution to the forces restoring nanotubes to their equilibrium state. Let us recall that individual SWCNT has four Goldstone degrees of freedom: rotation around its axis, translation as a whole along the tube axis and doubly degenerate translation as a whole in the direction normal to the axis. For individual SWCNTs comprising a system these degrees of freedom stop being of Goldstone type since neighboring nanotubes prevent free motion of each other. Thus in the system under consideration frequencies of all SWCNTs modes cannot vanish at $k\to 0$.

Let us note that elastic terms corresponding to restoring forces for translation and rotation of SWCNT as a whole should be significantly smaller (at least, one order less) than the term describing radial restoring force. Therefore we neglect tangential and rotational restoring forces and write free energy density of 2D membrane in the following simplest form:

$$g = \frac{\lambda}{2}(\varepsilon_{ii})^2 + \mu\varepsilon_{ij}^2 + K(\Delta H)^2 + \frac{C}{2}u_r^2, \qquad (1)$$

where $\lambda$ and $\mu$ are 2D analogs of Lame coefficients, $K$ is topological bending rigidity, $\Delta H = H - H_0$ with $H$ and $H_0$ standing for total and equilibrium mean curvatures of the surface, $C$ is pinning coefficient describing the radial homogeneous interaction between SWCNT and environment, $\varepsilon_{ij}$ is a 2D strain tensor which depends on $H_0$ [18] and $\boldsymbol{u} = (u_r, u_\varphi, u_z)$ is the 3D displacement field of the cylindrical membrane. The latter depends on the angle $\varphi$ and the variable $z$, which measures the distance along the cylinder axis. The curvature deviation linearized with respect to the field $\boldsymbol{u}$ and its derivatives reads $\Delta H = -(\Delta_s u_r)/(2R^2)$, where $\Delta_s = 1 + \partial_\varphi^2 + R^2\partial_z^2$.

Equations of motion for the 2D membrane are obtained by variation of the functional:

$$A = \int \left(g(\boldsymbol{u}) - \rho\dot{\boldsymbol{u}}^2/2\right)dSdt, \qquad (2)$$

where $t$ is time, $dS$ is the membrane area element, and $\rho$ is the surface mass density. To derive the motion equations we substitute $dS = Rdzd\varphi$, $\Delta H$, and $\varepsilon_{ij}$ in Eq. (2) and calculate the variation. The resulting equations have the following form:

$$\ddot{u}_r \rho R = -CRu_r - (\lambda + 2\mu)\left(\frac{u_r}{R} + \frac{\partial u_\varphi}{R\partial\varphi}\right) - \lambda\frac{\partial u_z}{\partial z} - K\frac{1}{R^3}\Delta_S^2 u_r,$$

$$\ddot{u}_\varphi \rho R = \frac{(\lambda + 2\mu)}{R}\left(\frac{\partial u_r}{\partial\varphi} + \frac{\partial^2 u_\varphi}{\partial\varphi^2}\right) + (\lambda + \mu)\frac{\partial^2 u_z}{\partial\varphi\partial z} + \mu R\frac{\partial^2 u_\varphi}{\partial z^2}, \qquad (3)$$

$$\ddot{u}_z \rho R = (\lambda + \mu)\frac{\partial^2 u_\varphi}{\partial\varphi\partial z} + (\lambda + 2\mu)R\frac{\partial^2 u_z}{\partial z^2} + \lambda\frac{\partial u_r}{\partial z} + \mu\frac{\partial^2 u_z}{R\partial\varphi^2}.$$

Substitution of $u_j = u_j^0 \exp(i(kz + n\varphi - \omega t))$, where $n$ is integer wave number, $k$ is one-dimensional wave vector, $\omega$ stands for circular frequency, and $j = r,\varphi,z$, yields dynamic matrix of the system (3). Vanishing of its determinant

$$\begin{bmatrix} \frac{\lambda+2\mu}{R} + \frac{KX^2}{R^3} + CR - R\rho\omega^2 & i\frac{(\lambda+2\mu)n}{R} & ik\lambda \\ -i\frac{(\lambda+2\mu)n}{R} & \frac{(\lambda+2\mu)n^2}{R} + \mu k^2 R - \rho\omega^2 R & (\lambda+\mu)nk \\ -ik\lambda & (\lambda+\mu)nk & (\lambda+2\mu)k^2 R + \frac{\mu n^2}{R} - R\rho\omega^2 \end{bmatrix}, \qquad (4)$$

where $X = (R^2k^2 + n^2 - 1)$, determines three real dispersion laws $\omega_j = \omega_j(k,n)$. Imaginary values of nondiagonal blocks in (4) reflect the $\pi/2$ phase shifts of the radial component with respect to the tangential ones.

Following [18], let us change the system of units to the experimental one, where CNTs dimensions are measured in nm, the length of wave vector $k$ is set in nm$^{-1}$, and solutions of det|**M**|=0 determine frequencies in cm$^{-1}$. We take the same estimations as in [18], in particular, $\lambda/\rho \approx 2400$ cm$^{-2}$ nm$^2$, $\mu/\rho \approx 5200$ cm$^{-2}$nm$^2$, $K/\rho \approx 12.5$ cm$^{-2}$ nm$^4$. The value of coefficient $C$ depends on material of CNTs environment. In order to demonstrate the influence of radial pinning on vibrational spectrum of SWCNT we assume that $C/\rho \approx 4000$ s$^{-2}$ (as an example) and plot dispersion curves for individual SWCNT (10, 10) and the same tube in medium (Fig. 1, solid black and dashed red curves, respectively).

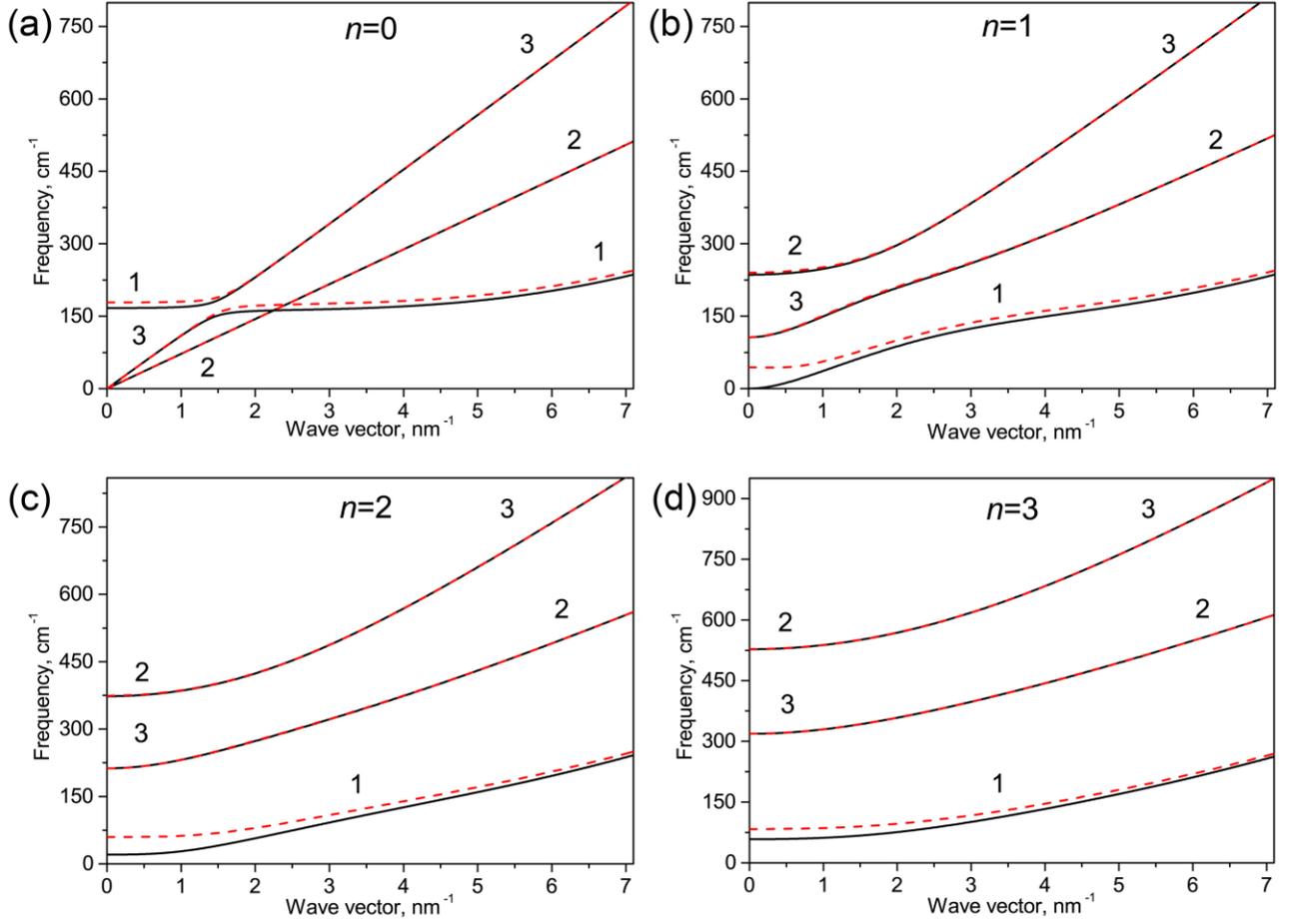

FIG.1. The dispersion curves for SWCNT (10, 10). Panels (a-d) represent the curves with $n=0..3$, respectively. Solid black curves correspond to individual nanotube and dashed red to one in a medium with $C/\rho \approx 4000$ s$^{-2}$. Let us note that all vibrations (except the special case $k=0$) are characterized by mixed polarization since nanotube surface is curved. Modes with mainly radial, tangential and longitudinal polarizations are denoted with numbers 1, 2 and 3, respectively. Principal direction of polarization may depend on the wave vector length. The interaction between modes of the same symmetry leads to resonance and change of principal direction of polarization: it occurs at $k\approx1$-2 nm$^{-1}$ ($n=0$ and $n=1$), $k\approx4$ nm$^{-1}$ ($n=2$) and $k\approx5$ нм$^{-1}$ ($n=3$).

Let us derive several particular solutions of equation det|**M**|=0 and obtain some analytical estimations for SWCNTs vibrational spectra. For $n=0$ and $k=0$ equality $M_{11}=0$ determines the dependence of RBM mode frequency on the nanotube radius and pinning coefficient $C$. As $\lambda R^2 \gg K$ and $\mu R^2 \gg K$ [18], we exclude the term $KX^2/R^3$ and write the expression for RBM mode frequency in the following form:

$$\omega_{RBM} = \sqrt{\frac{\lambda+2\mu}{R^2\rho} + \frac{C}{\rho}}. \quad (5)$$

Let us note that if $C/\rho<500$ s$^{-2}$ then the shift of RBM mode frequency is less than experimentally registrable instrumental error, namely 1 cm$^{-1}$.

For individual nanotube two modes with $n=0$ (modes 2 and 3 in Fig. 1a) linearly vanish with k→0 and correspond to transversal rotational and longitudinal acoustic modes. The slopes of these dispersion curves determine the sound velocities in SWCNT. In order to obtain analytically expressions for velocities of longitudinal ($V_{LA}$) and rotational ($V_{TA}$) modes one should find the derivatives of corresponding solutions of det|**M**|=0 with respect to $k$ assuming k→0, $n=0$, $C=0$. Taking into account $\lambda R^2 \gg K$ and $\mu R^2 \gg K$ [18] we obtain

$$V_{LA} = \sqrt{\frac{4\mu(\lambda+\mu)}{\rho(\lambda+2\mu)}}, \quad V_{TA} = \sqrt{\frac{\mu}{\rho}}. \quad (6)$$

Let us note that mode with $n=1$ (the first one in Fig. 1b) determines the main contribution to ULT heat capacity (in the third section we discuss this question in details). The mode vanishes quadratically and its dispersion law ($C=0$) at $k\to 0$ can be written as

$$\omega = \alpha k^2, \qquad (7)$$

where

$$\alpha = \frac{V_{LA} R}{\sqrt{2}}. \qquad (8)$$

For the nanotube in medium with account for radial pinning the frequencies of modes which propagate with velocities determined by Eq. (6) do not virtually shift, but the frequency of the mode described by Eq. (7) at $k\to 0$ increases by approximately $\sqrt{\frac{C}{2\rho}}$ (see the first mode in Fig. 1b). Let us note that taking into account tangential and rotational pinning leads to analogous upwards shifts at $k=0$ for corresponding acoustic modes. All these shifts dramatically decrease ULT heat capacity, as it is described in detail in the next section.

Before we calculate the heat capacity by means of obtained dispersion laws, let us clearly define the limitations of the continuous model we use. First, the SWCNTs phonons, which are indexed by $n$ and $k$ values, can be characterized by the effective wave vector

$$k_{eff} = \sqrt{k^2 + \left(\frac{n}{R}\right)^2}. \qquad (9)$$

Eq. (9), in turn, determines the effective wavelength $2\pi/k_{eff}$, which is to be several times longer than interatomic distance 0.142 nm in SWCNTs. Second, SWCNTs band structure is periodic with period $2\pi/a$ in reciprocal space. For the nanotube (10, 10) $2\pi/a \approx 25.5$ nm$^{-1}$. It means that at $k>12.25$ nm$^{-1}$ all dispersion curves on Fig. 1 should go down. Thus for the SWCNT under consideration the continuous model gives accurate results for $k$ less than 6-7 nm$^{-1}$. In this case the latter restriction is more severe than the one we can deduce from Eq. (8). Then, using Eq. (8) we derive that the maximum value for $n$: $n_{max}=5$. The obtained limitations to our continuous theory also constrain the temperature range of continuous models applicability for the heat capacity calculation.

### III. Discussion of SWCNTs low-temperature heat capacity and conclusions

In order to estimate temperature range of applicability for our approach we assume that modes with frequencies which do not satisfy the condition

$$\frac{h\nu}{k_B} \leq 9T$$

make negligibly small contribution to heat capacity. Here $h$ is Planck constant, $k_B$ stands for Boltzmann constant. The frequency of the lowest-energy mode (denoted as 1 in Fig. 1b) at wave vector length corresponding to the applicability limit of the continuous theory is approximately 200 cm$^{-1}$. At temperatures less than 37 K this mode does not practically excite as all the modes of higher frequencies. It is precisely this fact that determines the upper limit of the applicability range of our model as 37 K.

Now let us recall some of those properties of SWCNTs vibrational spectrum, which are determined by the nanotube symmetry. The factor group of SWCNTs with respect to translational subgroup is $D_N$ or $D_{Nh}$, where $N$ is even integer equal to the number of carbon hexagons contained within the nanotube unit cell [20]. For any SWCNT the modes with particular wave vector $k$ are indexed by integers $n = -N/2+1, -N/2+2...N/2$. Since the determinant (4) is quadratic in $n$ and $k$, the modes of the same absolute values of $n$ and $k$ are equivalent to each other. Thus the modes with nonzero $n$ and $k$ are fourfold degenerate. Besides, for every $n$ and $k$ there are 2x3=6 vibrational modes, where multiplier 2 corresponds to the number of atoms per graphene unit cell, and 3 stands for three possible polarizations of modes [21].

Let us denote the length of SWCNT as $L$ and its period as $a$. Thus the number of SWCNT atoms is given by $2NL/a$ and the total number of degrees of freedom is $6NL/a$. If the density of modes of particular type (for each $n$ value) in reciprocal space is denoted as $\delta$ then $6N\delta\frac{2\pi}{a} = 6N\frac{L}{a}$ and $\delta = \frac{L}{2\pi}$, since the number of normal vibrations should be equal to the number of degrees of freedom. Thus using Bose distribution [22] the total vibrational energy of SWCNT can be written in the following form:

$$U = \frac{L}{2\pi} \sum_{n=-\frac{N}{2}+1}^{n=\frac{N}{2}} \sum_{j=1}^{6} \int_{-\frac{\pi}{a}}^{\frac{\pi}{a}} \frac{h\nu_j(k,n)}{e^{\frac{h\nu_j(k,n)}{k_B T}} - 1} dk. \quad (10)$$

At low temperatures due to neglecting the high frequency modes, Eq. (10) can be simplified as:

$$U = \frac{L}{\pi} \sum_{-n_{\max}}^{n_{\max}} \sum_{j=1}^{3} \int_{0}^{\infty} \frac{h\nu_j(k,n)}{e^{\frac{h\nu_j(k,n)}{k_B T}} - 1} dk, \quad (11)$$

where $n_{\max}$ is the maximum number of oscillations obtained above. In Eq. (10) for each $j$ value 3 modes from 6 ones possess the high frequency since they are originated from the optic modes of graphene. In Eq. (11) these modes are eliminated from the sum.

Since the average diameter of SWCNTs in material under investigation [6] was equal to 1.25 nm, we simplify the problem and assume that the material consists of SWCNTs with this diameter only, as it was previously suggested in papers [9, 11-13, 16, 17]. Note, that the diameter 1.25 nm corresponds to the (10,10) nanotube, however, the continuous model does not use these indices. By differentiating Eq. (11) with respect to $T$ and numerical integrating the result with respect to $k$ we calculate the dependence of SWCNTs specific heat on temperature without account for radial pinning. Black squares show this dependence in Fig. 2, where the experimental data [6] in the range $2<T<37$ K are depicted by empty asterisks. The obtained theoretical curve approximates experimental data with an absolute error, which decreases only slightly with the temperature. Due to this fact in the LT range relative error of approximation takes a turn to the worse since specific heat decreases itself. Taking into account radial interaction between SWCNT and environment allows us to improve the correspondence between the theoretical and experimental curves significantly.

Indeed, there is some analogy with the well-known effect in superconductors: when upon the superconducting transition the gap in the electronic excitation spectrum opens the law describing the specific heat temperature behavior changes dramatically [23]. Fig.1 shows a similar change occurring in the lowest-energy phonon spectrum of individual SWCNT which interacts with the environment. The lowest branch with predominantly radial polarization and initially quadratic dispersion (see fig. 2b) demonstrates the upward frequency shift essential in the $k=0$ region. Unfortunately, the value of the pinning coefficient determining this change is hardly possible to be measured experimentally. Therefore we deduce its estimate from comparison of our calculation results with the experimental data. To do this in addition to the data [6] we also use the data on specific heat in the ULT region from Ref. 7. Let us denote the specific heat value measured at temperature $T_i$ as $C_{Vi}$ and deviation of the theoretical specific heat from the experimental one as $\Delta C_{Vi}$. By minimization of the mean-square relative error $\sum_i \left(\frac{\Delta C_{Vi}}{C_{Vi}}\right)^2$ we determine the effective pinning coefficient $C/\rho$. The curve corresponding to the specific heat with account for the obtained value $C/\rho=100$ s$^{-2}$ is plotted on Fig. 2 with red rhombuses.

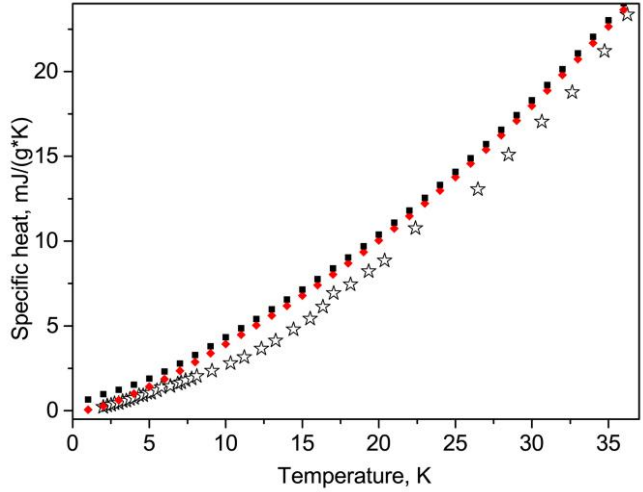

FIG. 2. Specific heat temperature dependences: theoretical ones with and without account for pinning (red rhombuses and black squares, respectively) and experimental one (empty asterisks) obtained from a sample consisting mainly of SWCNT bundles [6].

Let us stress, that all the models describing dynamics of *individual* SWCNTs are not suitable for calculation of ULT heat capacity of SWCNTs *bundle*. If in the ULT range ($T<2$ K) the interaction with environment is not taken into account then theoretical specific heat is *by order of magnitude greater* than experimental one. This statement concerns all earlier models [9-14]. Let us prove it. Obviously, at $T\rightarrow 0$ the main contribution to heat capacity comes from the vibrational mode with the quadratic dispersion (7). The existence of this mode and its type of dispersion at $k\rightarrow 0$ are universally recognized by all dynamic theories starting from the paper [11]. Preserving in (11) only a contribution from the mode with the dispersion (7) we obtain the following lower boundary for ULT heat capacity of SWCNTs:

$$C_V = \frac{3}{4}\zeta\left(\frac{3}{2}\right)\frac{Lk_B^2}{\sqrt{k_B \alpha h\pi}}\sqrt{T}, \qquad (12)$$

where $\zeta$ – zeta function (its value at 3/2 is approximately 2.61), $\alpha$ is determined by SWCNT dispersion law as a coefficient in Eq. (7), $h$ – Planck constant. If $\alpha$ is determined by Eq. (8) then the specific heat obtained from Eq. (12) at $T=1$ K is about 12 times greater than the experimental value [7]. In discrete model [12] parameter $\alpha$ is approximately a half of the value obtained according to Eq. (8). It means that at $T\rightarrow 0$ approach [12] is even worse than our model without account for radial pinning. Therefore, the paper [12] as the other ones [9-11, 13, 14] cannot explain experimental data in the ULT range, and their discussions of power $x$ in the $T^x$ law at $T<2$ K are pointless.

Only taking into account the interaction between SWCNT and environment allows understanding the heat capacity behavior in the ULT range. Obviously, the obtained value $C/\rho=100$ s$^{-2}$ is too small to be experimentally registered in the frequency shift of RBM mode, but appears to be sufficient to decrease ULT heat capacity of SWCNTs dramatically. Fig. 3 shows the ULT region, where the experimental data [7] fitted by phenomenological function $0.043T^{0.62}+0.035T^3$ [7] and the theoretical specific heat calculated using Eq. (11) at $C/\rho=100$ s$^{-2}$ are represented. For comparison Fig. 3 shows the lower boundary for ULT heat capacity of the free SWCNTs (not interacting with the environment).

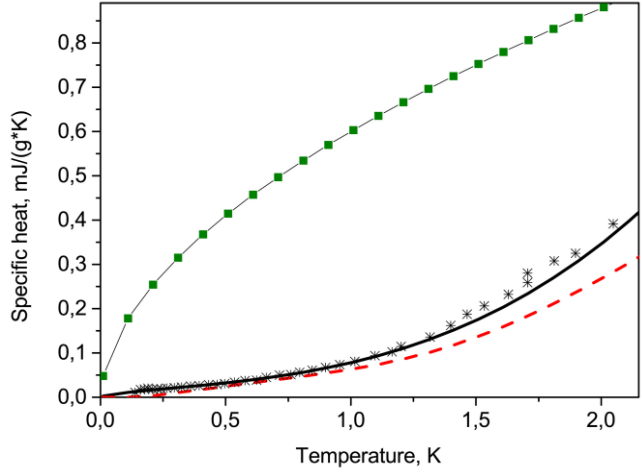

FIG. 3. Experimental data [7] fitted by approximating function $0.043T^{0.62}+0.035T^3$ [7] (black solid curve) and theoretical specific heat calculated in the frame of our model at $C/\rho=100$ s$^{-2}$ (red dashed curve). Green squares demonstrate function (12), where $\alpha$ is calculated from Eq. (8). All previous theoretical models [9-14] predict specific heat values above this curve.

In conclusion, we developed a continuous theory of SWCNTs dynamics with account for their weak interaction with environment. The obtained results are applied to calculation of SWCNTs specific heat in the LT and ULT ranges. We ascertain that ULT heat capacity cannot be correctly derived without account for environmental influence on SWCNT. The interaction between SWCNTs and medium makes additional contribution to the forces restoring nanotubes to their equilibrium state and significantly changes the dispersions of the phonon modes with frequencies vanishing at $k\rightarrow 0$ in individual SWCNT. Taking into account radial and tangential interaction between SWCNT and environment results in nonzero frequencies of these modes even at $k=0$. Radial interaction with medium appears to be more significant than tangential one and therefore is discussed in this paper in detail. We demonstrate that even weak radial interaction between SWCNT and environment leads to a sharp fall in phonon density of states at $\omega\rightarrow 0$ and in heat capacity at ULT. As far as we know, the approach proposed is the only one which results in theoretical specific heat dependence being in quantitative agreement with experimental data in the ULT range ($T<2$ K). Unfortunately, only two papers [7, 8] report results of measurements of ULT specific heat, and the dependences obtained differ significantly. This probably can be explained not only by inaccuracy of experiments but also by difference in pinning coefficient of the systems, which were studied. Thus the proposed theory can be refined by taking into account tangential interaction between SWCNTs and medium if new experimental data emerge.

It is possible that effective interaction between SWCNT and environment grows with temperature. If it will be shown in future, then, on the one hand, this fact will improve the agreement between theoretical and experimental data in the range $2<T<40$ K. On the other hand, the proposed theory or its modification can be applied to explain the shift of RBM mode frequency observed in [24]. And, last but not least, our theory can be explored to study LT specific heat of multi-walled CNTs and many other tubular structures (boron nitride, transition metal dichalcogenides) emerged in the past decade.

**Appendix. Discussion of some previous continuous models of SWCNTs phonon dynamics**

As far as we know the motion equations of cylindrical shell with thin walls were proposed 38 years ago [26] and the idea of considering SWCNTs as cylindrical shells of finite wall thickness was used later in a number of papers [15-17, 25]. Although a multitude of papers were devoted to SWCNTs dynamics, the equations [26], in our opinion, were familiar only to the authors of the approach [15]. We present these equations below in a slightly modified form and rename some quantities in order to avoid repetitions of notations: $\mu_P$ – Poisson's ratio, $E$ – Young's modulus, $\rho'$ – density of the material of the shell, $h$ – its thickness, $R$ – radius of equilibrium middle surface, $z$ and $\varphi$ – corresponding cylindrical coordinates, $\mathbf{u}=(u_r, u_\varphi, u_z)$ – 3D displacement field. Thus the motion equations [26] of the cylindrical shell are rewritten in the following form:

$$\frac{\mu_P}{R}\frac{\partial}{\partial z}u_z + \frac{1}{R^2}\frac{\partial}{\partial \varphi}u_\varphi + \left(\frac{1}{R^2} + \frac{h^2}{12}\left(\frac{\partial^2}{\partial z^2} + \frac{1}{R^2}\frac{\partial^2}{\partial \varphi^2}\right)^2\right)u_r = -\frac{\rho'(1-\mu_P^2)}{E}\frac{\partial^2 u_r}{\partial^2 t},$$

$$\frac{1+\mu_P}{2R}\frac{\partial^2}{\partial z \partial \varphi}u_z + \left(\frac{1-\mu_P}{2}\frac{\partial^2}{\partial z^2} + \frac{1}{R^2}\frac{\partial^2}{\partial \varphi^2}\right)u_\varphi + \frac{1}{R^2}\frac{\partial}{\partial \varphi}u_r = \frac{\rho'(1-\mu_P^2)}{E}\frac{\partial^2 u_\varphi}{\partial^2 t}, \quad (13)$$

$$\frac{\partial^2 u_z}{\partial z^2} + \frac{1-\mu_P}{2R^2}\frac{\partial^2}{\partial \varphi^2}u_z + \frac{1+\mu_P}{2R}\frac{\partial^2 u_\varphi}{\partial \varphi \partial z} + \frac{\mu_P}{R}\frac{\partial u_r}{\partial z} = \frac{\rho'(1-\mu_P^2)}{E}\frac{\partial^2 u_z}{\partial^2 t}.$$

Note that in the shell with the finite thickness the considerable part of the forces and their moments appear due to the strain tensor dependence on the distance from the middle surface [19, 27]. In addition, study [26] assumes that the radial displacement of the shell is much smaller than its macroscopic thickness and all the material constants are equal to the bulk ones.

Although SWCNTs have no macroscopic thickness $h$, it is possible to compare the systems of equations (13) and (3). For this purpose we exclude pinning term $CRu_r$ from the first equation of the system (3) and make the following substitution in (13):

$$\frac{E}{\rho'} = \frac{4}{\rho}\frac{\mu(\lambda+\mu)}{\lambda+2\mu}, \quad \frac{\mu_P}{\rho'} = \frac{1}{\rho}\frac{\lambda}{\lambda+2\mu}, \quad \frac{h}{\rho'} = \frac{2}{\rho}\sqrt{\frac{3K}{\lambda+2\mu}}. \quad (14)$$

We also rewrite operator $\left(\frac{\partial^2}{\partial z^2} + \frac{1}{R^2}\frac{\partial^2}{\partial \varphi^2}\right)^2$ (see the first equation in the system (13)) in easy-to-use form, namely $\left(\frac{\Delta'_s}{R^2}\right)^2$, where $\Delta'_s = \partial_\varphi^2 + R^2\partial_z^2$. The substitution (14) allows us to see a connection between material constants of the systems (3) and (13), including the relation between phenomenological bending rigidity $K$ of graphene sheet and macroscopic thickness $h$ of membrane.

The comparison between the systems under discussion shows that after the substitution (14) the second and the third equations of the system (13) become identical to those of the system (3). At the same time, the first equation of the system (13) differs from that of the system (3). Some terms containing higher derivatives with respect to variables $\varphi$ and $z$ are absent in (13). This absence is caused by the fact that the operator $\Delta_s = 1 + \partial_\varphi^2 + R^2\partial_z^2$ in the system (3) differs from the corresponding one $\Delta'_s$ in the system (13).

Note, that the absence of above mentioned terms in the first equation of the system (13) (caused by the missing unity in $\Delta'_s$) breaks the translational invariance of the system (13). Let us prove it. The uniform displacement $u_0$ of cylindrical shell as a whole in the plane perpendicular to its principal axis in the direction which is at an angle $\varphi_0$ with $e_x$ axis of Cartesian coordinate system reads

$$u_x = u_0 \cos(\varphi_0), u_y = u_0 \sin(\varphi_0), u_z = 0.$$

To express this displacement in the cylindrical coordinates, which are used in the system (13), we explore the following relations:

$$u_r = u_x \cos(\varphi) + u_y \sin(\varphi), \quad u_\varphi = -u_x \sin(\varphi) + u_y \cos(\varphi). \quad (15)$$

The displacement $u_z$ is obviously equal to zero. Substituting (15) in (13) we find that right side of the first equation does not vanish but becomes equal to $h^2 u_0 R^{-4}[\cos(\varphi)\cos(\varphi_0) + \sin(\varphi)\sin(\varphi_0)]/12$. According to the theory [26], which is used in paper [15], this fact implies that the Goldstone mode corresponding to the plane-parallel motion of a nanotube as a whole has nonzero frequency, but this inference is invalid for an individual (free) nanotube.

Finally, let us discuss the motion equations of cylindrical shell obtained in Ref. 16. Obviously, the authors of the paper [16] knew nothing about the motion equations from Ref. [26] and made an attempt to derive their own ones. They made a number of mathematically ill-founded assumptions resulting in equations, where several terms were missing with respect to the system (13). Note that equations from the paper [16] are easily obtained from the system (13) by vanishing of *h* and excluding the term $\frac{1}{R^2}\frac{\partial u_r}{\partial \varphi}$ from the second equation of the system (13). Let us stress that according to the relations (14), vanishing of the membrane macroscopic thickness *h* means that the membrane has no bending rigidity *K*. This assumption is obviously incorrect and leads to invalid results. In particular, according to this approach, for every *n* value there exists a mode with nonzero frequency at *k*=0. Besides, the system of equations obtained in the paper [16] also assigns nonzero frequency to the mode corresponding to the plane-parallel motion of a nanotube as a whole. By substitution of the uniform displacement (15) in this system we obtain that only its first and second equations vanish identically. The right side of the third equation becomes equal to $u_0 R^{-2}[\cos(\varphi_0)\sin(\varphi) - \sin(\varphi_0)\cos(\varphi)]$, and it is obviously caused by the absence of the term $\frac{1}{R^2}\frac{\partial u_r}{\partial \varphi}$ in the left side. Also note that the right side of the first equation vanishes identically only due to the absence of the terms related to the membrane bending rigidity in the approach [16].

**Acknowledgements**

This work was supported by the RFBR grant 13-02-12085 ofi_m.

**References**
[1] S. Iijima, Nature **354**, 56–58 (1991).
[2] R. Saito, G. Dresselhaus and M.S. Dresselhaus, *Physical Properties of Carbon Nanotubes* (Imperial College Press, London, 1998).
[3] S. Reich, C. Thomsen and P. Ordejon, *Elastic Properties and Pressure-induced Phase Transitions of Single-walled Carbon Nanotubes*, Vol. **235** (Wiley Online Library, 2003).
[4] A. Jorio, M.S. Dresselhaus and G. Dresselhaus, *Carbon Nanotubes: Advanced Topics in the Synthesis, Structure, Properties and Applications*, Vol. **111** (Springer, Berlin, 2008).
[5] A. Mizel *et al.*, Physical Review B **60**, 3264-7 (1999).
[6] J. Hone, B. Batlogg, Z. Benes, A. T. Johnson and J. E. Fischer, Science **289**, 1730-3 (2000).
[7] J. C. Lasjaunias, K. Biljaković, Z. Benes, J. E. Fischer, and P. Monceau, Physical Review B **65**, 113409-4 (2002).
[8] B. Xiang, C.B. Tsai, C.J. Lee, D.P. Yu, Y.Y. Chen, Solid State Communications **138**, 516–520 (2006).
[9] E. Dobardžić, I. Milošević, B. Nikolić, T. Vuković, and M. Damnjanović, Physical Review B **68**, 045408-9 (2003).
[10] J. X. Cao, X. H. Yan, Y. Xiao, Y. Tang, and J. W. Ding, Physical Review B **67**, 045413-6 (2003).
[11] V. N. Popov, Carbon **42**, 991-995 (2004).
[12] J. Zimmermann, P. Pavone, and G. Cuniberti, Physical Review B **78**, 045410-13 (2008).
[13] Michael C H Wu and Jang-Yu Hsu, Nanotechnology **20,** 145401-6, (2009).
[14] Chunyu Li and Tsu-Wei Chou, Physical Review B **71**, 075409-6 (2005).
[15] S. S. Savinskiĩ and V. A. Petrovskiĩ, Physics of the Solid State **44**, No. 9, 1802–1807 (2002).
[16] S. Zhang, M. Xia, S. Zhao, T. Xu and E. Zhang, Physical Review B **68**, 075415-7 (2003).

[17] S.P. Hepplestone, A.M. Ciavarella, C. Janke, G.P. Srivastava, Surface Science **600**, 3633–3636, (2006).
[18] S. B. Rochal, V. L. Lorman and Yu. I. Yuzyuk, Physical Review B **88**, 235435-6 (2013).
[19] L. D. Landau and E. M. Lifshitz, *Theory of Elasticity* (Addison-Wesley, Reading, MA, 1959).
[20] O. E. Alon, Phys. Rev. B **63**, 201403 (2001).
[21] M. V. Avramenko, S. B. Rochal, and Yu. I. Yuzyuk, Phys. Rev. B **87**, 035407 (2013).
[22] Ch. Kittel, *Introduction to Solid State Physics* (Wiley, USA, 2005).
[23] N. W. Ashcroft, N. D. Mermin, *Solid State Physics* (Holt, Rinehart and Winston, USA, 1976).
[24] R. Saito, M. Hofmann, G. Dresselhaus, A. Jorio, and M. S. Dresselhaus, Advances in Physics **60** (3), 413–550 (2011).
[25] G.D. Mahan, Phys. Rev. B **65**, 235402 (2002).
[26] A. S. Volmir, *Shells in Liquid and Gas Flow* (Nauka, Moscow, 1976, in Russian).
[27] A. E. H. Love, *A Treatise on the Mathematical Theory of Elasticity* (Cambridge, at the University Press, 1927).